%% file: Main.tex
\UseRawInputEncoding
\documentclass[%reprint,
superscriptaddress,
bibnotes,
 amsmath,amssymb,
 aps,
pra,
twocolumn,
longbibliography
]{revtex4-1}
\usepackage{graphicx}% Include figure files
\usepackage{dcolumn}% Align table columns on decimal point
\usepackage{xr}
\usepackage{bm} %bold math
\usepackage{float}
\usepackage{hyperref} %add hyperref capabilities
\hypersetup{
	colorlinks=true,
	linkcolor=blue,
	citecolor=blue,
	filecolor=magenta,
	urlcolor=blue         
}
\usepackage{color,soul}
\usepackage{lineno} %linenumbers

\renewcommand{\selectlanguage}[1]{}

\begin{document}

\title{Multi-phonon Fock state heralding with single-photon detection}

\author{Andrew J. Shepherd}
\email[]{as5262@nau.edu}
\affiliation{Department of Applied Physics and Materials Science, Northern Arizona University, Flagstaff, AZ 86011, USA}
\affiliation{Center for Materials Interfaces in Research and Applications (¡MIRA!), Flagstaff, AZ, USA}

\author{Ryan O. Behunin}
\email[]{ryan.behunin@nau.edu}
\affiliation{Department of Applied Physics and Materials Science, Northern Arizona University, Flagstaff, AZ 86011, USA}
\affiliation{Center for Materials Interfaces in Research and Applications (¡MIRA!), Flagstaff, AZ, USA}

\date{\today}

\begin{abstract}
Recognized as a potential resource for quantum technologies and a possible testbed for fundamental physics, the control and preparation of nonclassical states of mechanical oscillators has been explored extensively. 
Within optomechanics, quantum state synthesis can be realized by entangling photonic and phononic degrees of freedom followed by optical detection. Single-photon detection enables one of the most powerful forms of such heralded quantum state preparation, permitting the creation of single phonon states when applied to conventional cavity optomechanical systems. 
As the complexity of optomechanical systems increases, single-photon detection may provide heralded access to a larger class of exotic quantum states. Here, we examine the quantum dynamics of optomechanical systems that permit forward Brillouin scattering, where a single phonon mode mediates transitions between a collection of equally spaced optical resonances. Solving both the Schrodinger equation and the Lindblad master equation for this system, we find that initial states comprised of single photons or weak laser pulses evolve into complex quantum states where the frequency of single photon states and the phonon occupation number are entangled. Physically, these interactions permit a single photon to scatter to lower frequencies, where phonon excitation occurs for each scattering event. Combining this result with frequency filtering, we show how single-photon detection can herald selected multi-phonon Fock states, even in the presence of optical losses. We also present an approach for quantum tomography of the heralded phonon states.
\end{abstract}

\maketitle

Heralding techniques have proven to be a powerful tool for manipulating entangled quantum states. By measuring a component of an entangled system, it is possible to collapse the remainder of the system into a known quantum state. Single-photon detection is used extensively for these purposes, enabling, for example, the production of single photons via parametric down-conversion \cite{couteau_spontaneous_2018, caspani_integrated_2017} and the preparation of photon-added coherent states \cite{zavatta_quantum--classical_2004}. 
Within the field of cavity optomechanics, heralded state preparation through single-photon detection has permitted the realization of single phonon states \cite{hong_hanbury_2017, galland_heralded_2014}. 
Due to potentially long quantum coherence times \cite{goryachev_extremely_2012, goryachev_observation_2013, maccabe_nano-acoustic_2020, k_j_satzinger_quantum_2018, chu_creation_2018} and large effective mass, such mechanical oscillators are promising candidates for applications in quantum information science, such as transducers \cite{stannigel_optomechanical_2010, jiang_efficient_2020}, memories \cite{pechal_superconducting_2018}, repeaters \cite{pechal_superconducting_2018} and sensors \cite{mason_continuous_2019},  as well as a test-bed of fundamental physics \cite{marshall_towards_2003, diosi_models_1989, penrose_gravitization_2014, ghirardi_markov_1990}. 
However, advancing the applications mentioned above necessitates the development of systems and protocols that can access a wider variety of nonclassical phonon states \cite{chu_creation_2018,chu_quantum_2017,k_j_satzinger_quantum_2018,behunin_harnessing_2023}. 
\indent Here, we examine the quantum dynamics of optomechanical systems that utilize \textit{forward} Brillouin interactions.
In sharp contrast with conventional backward Brillouin scattering \cite{behunin_harnessing_2023}, a single phonon mode mediates interactions between neighboring resonances in a ladder of optical modes \cite{shelby_resolved_1985, raymer_stimulated_1981, kharel_noise_2016}. These dynamics are made possible in highly confined waveguides (or resonators), where the participating slow group velocity mechanical modes have ($\sim$MHz-GHz) frequencies determined by the system geometry.    
Taking the phonon to occupy the ground state initially, we solve the Schrodinger equation for a single photon input, as well as for a weak laser input, i.e., small coherent state amplitude. We find that the scattering of the injected photon to lower frequencies corresponds to an increasing phonon occupation number, generating entangled states where the mode of single photon occupation is correlated with the phonon occupation number. When combining this entanglement with single-photon detection of one of the resonant optical modes, our results present a path to heralded multi-phonon Fock states.

{\it Forward Brillouin Scattering}: Brillouin scattering involves inelastic light-matter interactions, where incident light will shift frequency upon interaction with acoustic phonons. The more commonly known backward Brillouin scattering involves two counter-propagating light waves that can be coupled with a travelling sound wave via electrostriction \cite{boyd_chapter_2008}. Confined systems, such as waveguides and resonators, permit a form of forward Brillouin scattering (FBS) where a cut-off mechanical mode can transfer energy between copropagating optical resonances \cite{shelby_resolved_1985}. 

In FBS, this phonon mode with frequency $\Omega$ mediates scattering between a ladder of optical modes with frequencies labeled $\omega_m$, where $m$ is a general integer placeholder labeling each optical mode. This process is most efficient when the participating optical and mechanical waves satisfy phase-matching conditions, akin to energy and momentum conservation, given by 
\begin{align}
\label{eq:phase-matching-1}
    \omega_m = \omega_{m-1} + \Omega \\
\label{eq:phase-matching-2}
    k_m = k_{m-1} + q, 
\end{align}

\begin{figure}
    \centering
    \includegraphics[width=8.6cm]{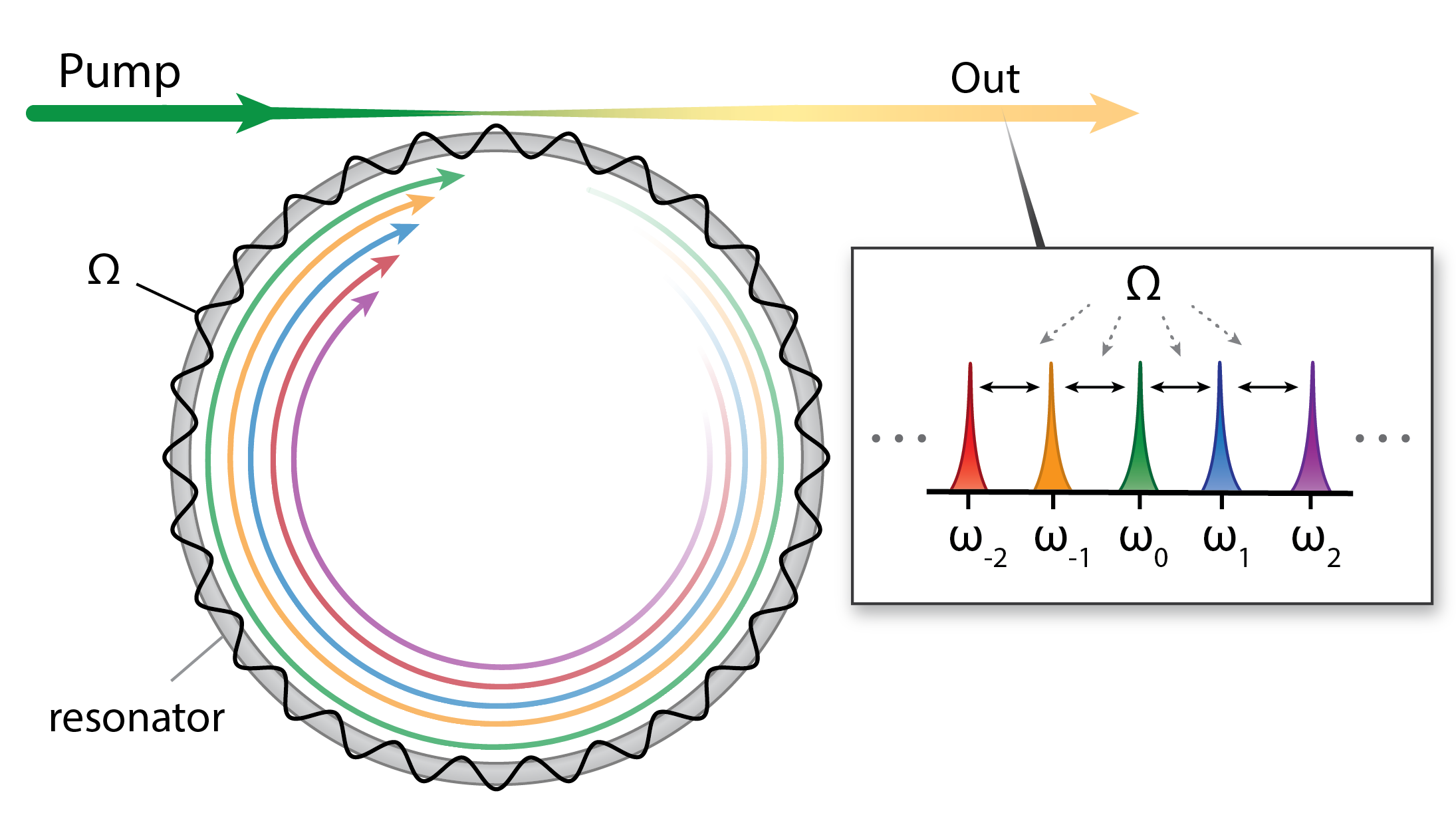}
    \caption{Proposed physical implementation of a FBS system, where a phonon, $\Omega$, mediates transitions between a ladder of optical resonances in an optical resonator.}
    \label{fig:resonator}   
\end{figure}
\begin{figure}
    \centering    
    \includegraphics[width=8.6cm]{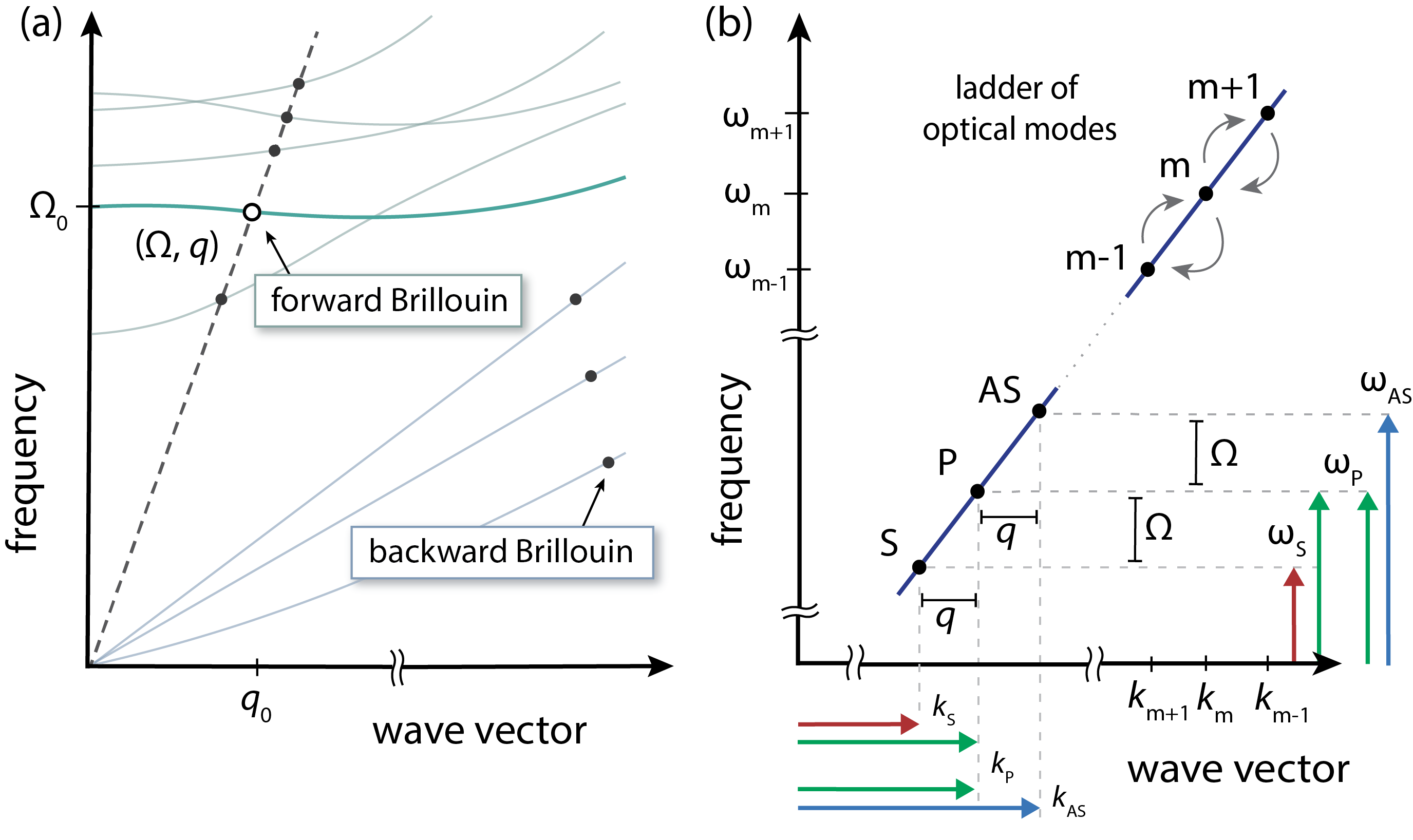}
    \caption{Phase matching conditions for intra-modal forward Brillouin scattering (in a waveguide). (a) Acoustic dispersion, with the forward Brillouin active mode shown as a white circle, where the mechanical frequency is approximately equal to the cut-off frequency, i.e. $\Omega \cong \Omega_0$. Other Brillouin active modes, including backward Brillouin, shown with black dots. (b) Optical dispersion, where the phonon frequency and wave vector resonantly couples adjacent optical modes.}
    \label{fig:dispersion}   
\end{figure}

 
\noindent
where $k_m$ and $q$ are the wavevectors for the $m$th optical mode and phonon respectively. Using the optical dispersion relation $\omega_m = c k_m/n_m$, where $n_m$ is the effective index evaluated at $\omega_m$, $c$ is the speed of light, and noting that $k_{m-1} \approx k_m - (\omega_m - \omega_{m-1}) \partial k_m/\partial \omega$, Eqs. \eqref{eq:phase-matching-1} and \eqref{eq:phase-matching-2} require that the phonon phase velocity is equal to the optical group velocity. While the phase velocity of sound is much less than light in bulk materials, Eqs. \eqref{eq:phase-matching-1} and \eqref{eq:phase-matching-2} can be satsified in confined systems that support cut-off mechanical modes that can oscillate without propagating, i.e., $\Omega \neq 0$ when $q \approx 0$, (see Fig. \ref{fig:resonator} and \ref{fig:dispersion}).
The Brillouin active mechanical mode is denoted in Fig. \ref{fig:dispersion}(a) as the white circle on the cut-off phonon dispersion curve. The optical dispersion and phase matching for these modes is shown in Fig. \ref{fig:dispersion}(b). While scattering to distant optical modes is physically limited by the optical dispersion, these effects are negligible for a broad variety of candidate systems over the relatively narrow frequency ranges of FBS. For example, recent experiments demonstrate cascading to over 25 Stokes orders  \cite{shanavas_cascaded_2022}. It is important to note that the dispersion curves in Fig. \ref{fig:dispersion} hold for waveguide systems, and differences may occur when looking at other FBS-capable systems, e.g., discretization of allowable wavevectors in resonators.

In contrast with more familiar forms of Brillouin light scattering, in FBS the phonon frequency and optomechanical coupling are highly tailorable with changes in the waveguide geometry, leading to selective resonant phonon frequencies and coupling strengths \cite{kharel_noise_2016, rakich_giant_2012}. These properties have been explored in a broad variety of systems \cite{eggleton_brillouin_2019}, including; telecommunications  \cite{shelby_resolved_1985}, photonic crystal \cite{kang_tightly_2009, kang_all-optical_2010, renninger_guided-wave_2016, renninger_forward_2016}, and liquid-core  \cite{behunin_spontaneous_2019} fibers; silicon waveguides \cite{shin_tailorable_2013, kittlaus_large_2016, shin_control_2015}; microring \cite{zhang_forward_2017} and microfluidic resonators \cite{bahl_brillouin_2013}; and optical microspheres \cite{bahl_stimulated_2011, yu_investigation_2022}. Related systems comprising optical resonators that contain highly confined mechanical waveguides are suitable to realize FBS optomechanics when the mechanical frequency $\Omega$ is an integer multiple of the optical free-spectral range (e.g., see Fig. \ref{fig:resonator}).

{\it Quantum Dynamics}: To model the quantum dynamics of an FBS optomechanical system, we use the Hamiltonian $H = H_0 + H_{int}$,
\begin{align}
    \label{eq:H0}
    H_0 = \displaystyle\sum^\infty_{m=-\infty}\hbar (\omega_p + m\Omega)a^{\dag}_m a_m + \hbar \Omega b^{\dag}b \\
       \label{eq:Hint}
    H_{int} = \hbar g \displaystyle\sum^\infty_{m=-\infty} a_m a^{\dag}_{m-1}b^{\dag} + H.c. 
\end{align}
\noindent 
where $H_0$ and $H_{int}$ describe the free evolution and interactions of the system respectively. Here, $\Omega$, $b$ and $b^\dag$ are the angular frequency, annihilation operator and creation operator for the phonon mode.  With $\omega_p$ as a central resonant frequency, the first term in $H_0$ describes the collection of resonantly spaced optical modes, where $a_m$ is the annihilation operator for the $m^{th}$ optical mode that has the angular frequency $\omega_m = \omega_p + m\Omega$. As a consequence of the definition of $\omega_m$, ideal phase matching is assumed throughout, i.e., $\omega_m = \omega_{m-1} +  \Omega$. 
To reflect the independence of these distinct optical modes in the absence of FBS, the annihilation and creation operators satisfy $[a_m,a^\dag_{m'}] = \delta_{mm'}$ where $\delta_{mm'}$ is the Kronecker delta. The interaction Hamiltonian $H_{int}$, and associated constant coupling rate $g$, quantify the strength of forward Brillouin interactions which can be produced by electrostriction and radiation pressure \cite{kharel_noise_2016, rakich_giant_2012}. 

By considering dynamics occurring at times much shorter than reported phonon decoherence times, we ignore loss on the mechanical mode. This regime can be accessed when the Brillouin coupling rate exceeds the mechanical decoherence rate. As one example, forward Brillouin coupling rates as high as $g \sim (2\pi) 29$ kHz and mechanical decoherence rates as low as $T_2^{-1} \sim (2\pi) 1.2$ kHz have been realized silicon systems \cite{shin_tailorable_2013, kittlaus_large_2016, maccabe_nano-acoustic_2020}. The impact of optical losses is neglected to begin but later analyzed by solving the Lindblad master equation. Other than decreasing the probability of a heralding event, optical dissipation does not alter the essential conclusions of the lossless model.
 
Neglecting decoherence, an analytical solution of the system's dynamics can be found through application of the time evolution operator on an initial state: $|\psi(t)\rangle = U(t,0)|\psi(0)\rangle$, with $U = \exp\{-iHt/\hbar\}$. Under the condition of phase-matched optical and mechanical modes, the FBS interaction described by $H_{int}$ conserves energy (i.e.,  $[H_0,H_{int}]=0$), permitting the time evolution operator to be factorized in the following form
\begin{equation}
\label{eq:time-evolution-operator}
    U(t,0) = U_0 U_{int}
\end{equation}
with $U_0=\exp\{-iH_0t/\hbar\}$ and $U_{int}=\exp\{-iH_{int}t/\hbar\}$. Taking advantage of the infinite discrete symmetry of the system (i.e., the involvement of an infinite ladder of equally spaced optical modes and constant coupling $g$ in Eqs. \eqref{eq:H0} and \eqref{eq:Hint}), $U_{int}$ can be factored further using the Glauber formula, a special case of the Zassenhaus expansion,
\begin{equation}
    e^{X+Y} = e^{X}e^{Y}e^{-\frac{1}{2}[X,Y]}
\end{equation}
where $X$ and $Y$ are placeholders for any operators satisfying $[X,[X,Y]] = 0$ and $[Y,[X,Y]] = 0$ \cite{dupays_closed_2023}. 
For $U_{int}$, the operators $X$ and $Y$ are given by
\begin{align}
    X = -i g t A b^{\dag} \quad {\rm and} \quad
    Y = -i g t A^\dag b 
\end{align}
where 
\begin{align}
    A = \displaystyle\sum^\infty_{m=-\infty} a_m a^{\dag}_{m-1} \quad {\rm and} \quad A^\dag = \displaystyle\sum^\infty_{m=-\infty} a^{\dag}_m a_{m-1}.
\end{align}
Noting that $A$ and $A^\dag$ commute 
\begin{align}
    [A,A^\dag] & =  \displaystyle\sum^\infty_{m=-\infty} \displaystyle\sum^\infty_{m'=-\infty} [a_m a^{\dag}_{m-1}, a^{\dag}_{m'} a_{m'-1}] \\ & = \displaystyle\sum^\infty_{m=-\infty} (n_{m-1}-n_m) = 0, \nonumber
\end{align}
we find $[X,Y] = g^2 t^2 A A^\dag$ and $[X,[X,Y]] = [Y,[X,Y]]=0$, permitting application of the Glauber formula and leading to the factorized form of $U_{int}$ given by
\begin{equation}
\label{eq:Uint}
\begin{split}
     U_{int} = &e^{-igt A b^\dag} e^{-igtA^\dag b} e^{-\frac{1}{2}g^2t^2 A A^\dag}.
\end{split}
\end{equation}

Assuming that the phonon is initially found in the ground state and that a single photon is injected into the $m=0$ mode, the initial state is $|\psi(0)\rangle = |...,0_{-1},1_0,0_1,...\rangle \otimes|0\rangle_{ph}$. Here, the first ket describes the ladder of optical modes with each mode separated by a comma and where the suffix labels the optical mode. The second ket describes the state of the phonon, i.e, $|0\rangle_{ph}$. Explicit calculation of $U(t,0)|\psi(0)\rangle$ gives 
\begin{equation}
\label{eq:wavefunction}
    \begin{split}
    |\psi(t) \rangle =&e^{-i\omega_0t}e^{-\frac{g^2t^2}{2}}\displaystyle\sum^{\infty}_{n=0}\frac{(-igt)^n}{\sqrt{n!}} |\varphi_n\rangle\otimes |n\rangle_{ph}
    \end{split}
\end{equation}
where we have introduced the compact notation 
\begin{align}
    |\varphi_n\rangle = |...0_{-n-1},1_{-n},0_{-n+1},...\rangle.
\end{align}
This entangled superposition state quantifies how the input photon is Stokes shifted in frequency down the resonant optical ladder while simultaneously increasing the occupation number of the phonon mode. 
\begin{figure}
    \centering  
    \includegraphics[width=8.6cm]{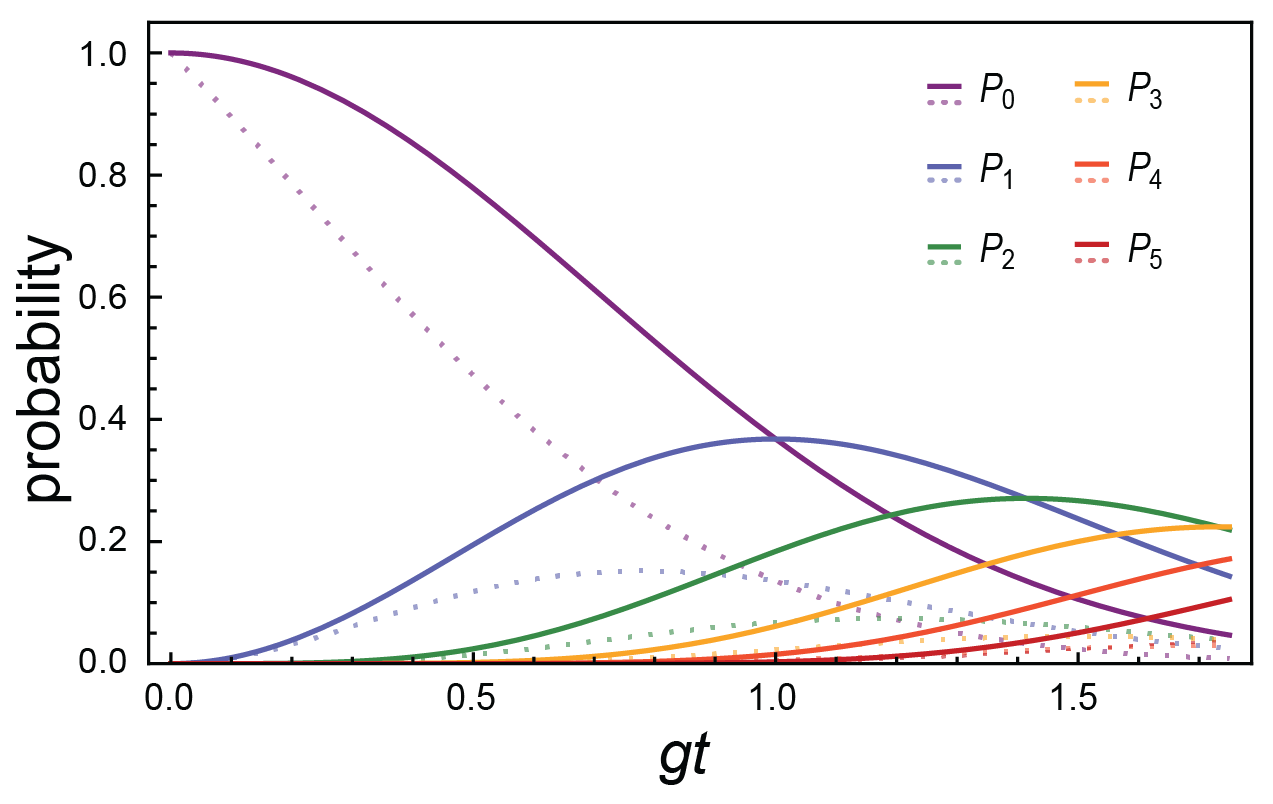}
    \caption{Probabilities of the first few accessible states when FBS system is stimulated with a single photon, plotted over $gt$ (g as the coupling rate, t as time). Solid lines hold for no optical loss, ($\gamma=0$). Dotted lines include optical losses, with $\gamma=g$.}
   \label{fig:probabilities}
\end{figure}

\begin{figure}
    \centering
    \includegraphics[width=8.6cm]{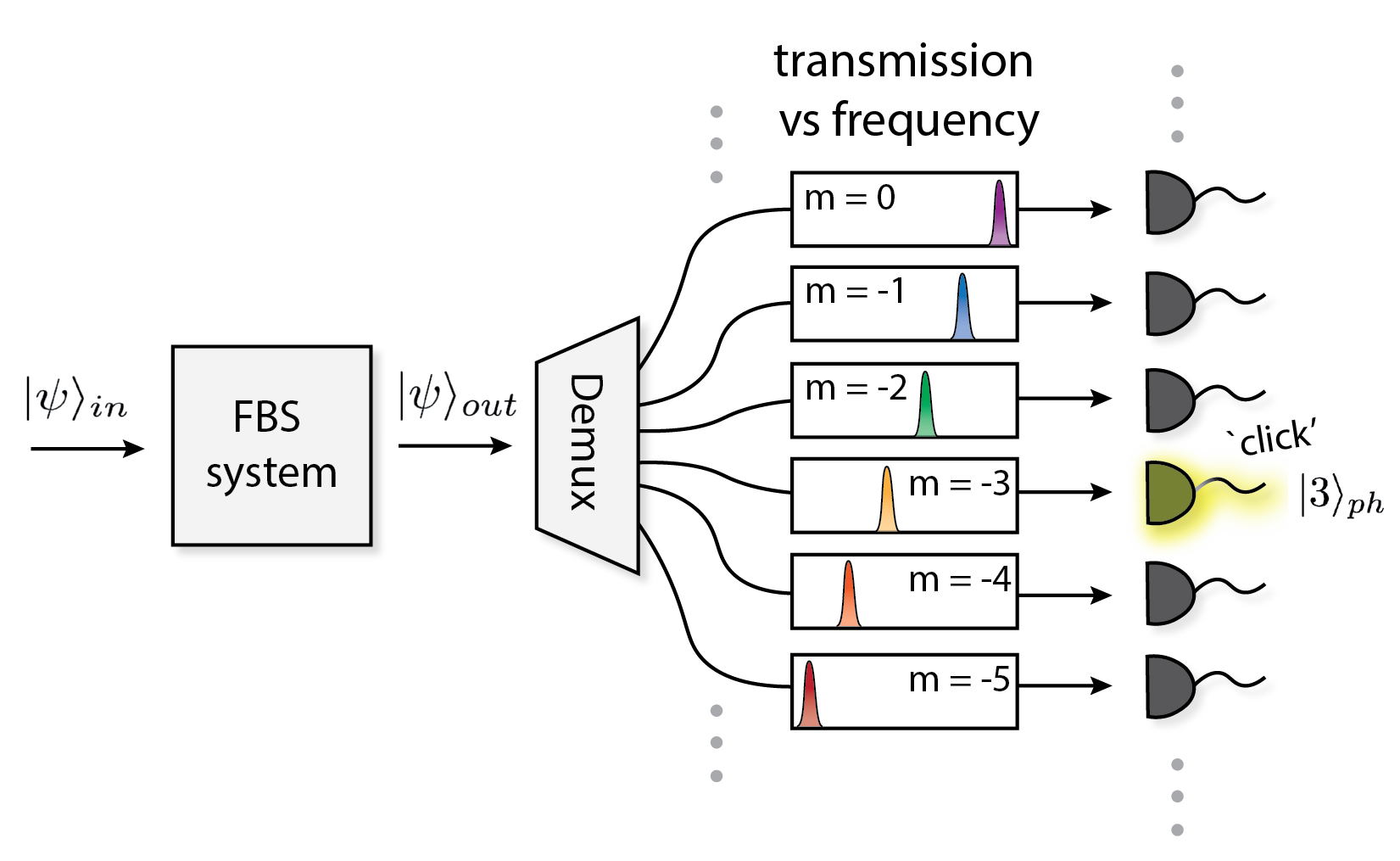}
    \caption{Scheme describing quantum state synthesis in a forward Brillouin scattering system. The entangled state, $|\psi\rangle_{out}$, goes through a demultiplexer, filtering by frequency into different transmission lines to an array of single-photon detectors. The detection event (shown by the yellow highlight) projects the phonon into a Fock state.}
    \label{fig:FBS-Scheme}  
\end{figure}

Equation \eqref{eq:wavefunction} shows how measuring a photon of a specified frequency $\omega_j$ can collapse the wave-function and isolate a highly non-classical phonon quantum state. To achieve this, the quantum state is spatially entangled by demultiplexing the emitted optical state as a function of frequency, where subsequent single-photon detection in the $-j\textsuperscript{th}$ optical mode signals the collapse of the wavefunction into a phonon Fock state of known occupation number. The probability for this heralding event  to occur is given by
\begin{align}
\label{eq:probability}
    P_j = |(\langle \varphi_j | \otimes \langle j |_{ph})| \psi(t)\rangle|^2.
\end{align}
Fig. \ref{fig:probabilities} plots Eq. \eqref{eq:probability} as solid lines for the first few accessible states. An example of this heralding process is shown in Fig. \ref{fig:FBS-Scheme}, where $|\psi\rangle_{in}$ is the injected photon and $|\psi\rangle_{out}$ is the entangled state to be filtered by frequency into different transmission lines, leading to single-photon detectors. In this example, the detected photon has been Stokes shifted to mode $m=-3$, corresponding to the generation of an $n=3$ phonon Fock state.

Although permitting analytical simplicity, single photon inputs present greater experimental challenges, requiring a heralding process to achieve the desired initial state. A more physically accessible initial state would be the input of a weak coherent state (faint laser), where the initial state of the $m=0$ optical mode is given by $|\alpha \rangle \approx |0\rangle + \alpha|1\rangle$ ($|\alpha| \ll 1$), taking the place of $|\psi\rangle_{in}$ in Fig. \ref{fig:FBS-Scheme}. Applying the time evolution operator on $|\psi(0)\rangle \approx |vac\rangle\otimes |0\rangle_{ph} + \alpha|\varphi_0\rangle \otimes |0\rangle_{ph}$ gives the wavefunction at time $t$
\begin{align}
\label{eq:psi-weak-laser}
    |\psi &(t)\rangle \approx  |vac\rangle \otimes |0\rangle_{ph} \\ &+ \alpha e^{-i\omega_pt}e^{-\frac{g^2t^2}{2}}\displaystyle\sum_{n=0}^\infty\frac{(-igt)^n}{\sqrt{n!}}|\varphi_n\rangle \otimes |n\rangle_{ph} \nonumber
\end{align}
with $|vac\rangle$ denoting vacuum overall optical modes. For this wavefunction, the vacuum state dominates in probability; however, the phonon manipulation relies on conditional measurement. While the detection probability is low, the entire state collapses upon detection, heralding a phonon Fock state in the same manner as before. 

We analyze the effects of optical losses using the Lindblad master equation. When weakly coupled to a zero-temperature Markovian bath, the density matrix for the system satisfies 
\begin{align}
\label{eq:Lindblad}
    \dot{\rho} = -\frac{i}{\hbar} [H, \rho] + \frac{\gamma}{2} \sum_{m=-\infty}^\infty (
    2 a_m\rho a_m^\dag - a^\dag_m a_m \rho - \rho a^\dag_m a_m
    ),
\end{align}
producing exponential decay for each optical mode at the decay rate $\gamma$ \cite{meystre_quantum_2021}. Restricting our attention to single photon input and accounting for losses modeled by Eq. \eqref{eq:Lindblad}, the density matrix can be represented by 
\begin{align}
\label{eq:density-matrix}
\rho = \sum_{n,n'} \bigg[
\alpha_{n,n'}|\varphi_n\rangle \langle \varphi_{n'} |
&+ \beta_{n,n'}\delta_{nn'} 
|vac\rangle \langle vac |
\bigg] \nonumber \\&\otimes |n\rangle_{ph} \langle n' |_{ph}
\end{align}
where Eq. \eqref{eq:Lindblad} provides the following equations for the time-dependent coefficients $\alpha_{n,n'}$ and $\beta_{n,n'}$ 

\begin{align}
  \dot{\alpha}_{n,n'}  = & 
  -ig(\sqrt{n}\alpha_{n-1,n'} +\sqrt{n+1}\alpha_{n+1,n'} \nonumber \\
 & -\sqrt{n'}\alpha_{n,n'-1} -\sqrt{n'+1}\alpha_{n,n'+1})
  -\gamma \alpha_{n,n'} 
  \\
  \dot{\beta}_{n,n'} = & \gamma \delta_{nn'} \alpha_{n,n'}.
\end{align}
The solution to these two equations, given by
\begin{align}
    \alpha_{n,n'} & = \frac{(-i g t)^n (i g t)^{n'}}{\sqrt{n!} \sqrt{n'!}}e^{-g^2 t^2 -\gamma t} 
    \\
    \beta_{n,n'} & = \delta_{n,n'}\gamma \int_{0}^t d\tau \frac{(g \tau)^{2n}}{n!}e^{-g^2 \tau^2 -\gamma \tau}
\end{align}
can be shown to preserve unitarity (i.e., ${\rm tr}(\rho) =1$), and give the probability of single photon detection in the $-j\textsuperscript{th}$ mode and collapse to a phonon Fock state of number $j$, 
\begin{align}
   \langle \varphi_j |  \rho | \varphi_j \rangle  =  \alpha_{jj}  |j\rangle_{ph} \langle j|_{ph}
    = P_j e^{-\gamma t} |j\rangle_{ph} \langle j|_{ph} .
\end{align}
This result shows that although the detection probability is suppressed by $\exp\{ - \gamma t\}$, the entanglement between single photon occupation and phonon Fock state is preserved. Thus, optical losses do not fundamentally alter the conclusions of our analysis.

After the detection event, readout and manipulation of the phonon state can be done through optomechanical $\pi$- and $\pi/2$-pulses \cite{behunin_harnessing_2023}, where a beam-splitter interaction between phonons and photons can be realized by intense optical pulses. In an FBS system, enabling this requires an engineered `stop-band' on a resonant mode of higher frequency than the input photon, where the occupation is forbidden. Otherwise, an intense optical pulse will simultaneously lead to beam-splitting and two-mode squeezing interaction. Leveraging the semi-compact Hilbert space of Eq. \eqref{eq:wavefunction}, where the injected photon cannot scatter to frequencies higher than $m=0$, Eq. \eqref{eq:wavefunction} satisfies the Schrodinger equation with a single photon input in the presence of a stop-band on any mode $m=j>0$. Suppression of mode $\omega_j$ removes the terms containing $a_j$ or $a_j^\dag$ from the sum in Eqs. \eqref{eq:H0} and \eqref{eq:Hint}, leaving only an anti-Stokes sideband for the $\omega_{j+1}$-mode. Thus, strong pulses with frequency $\omega_{j+1}$ can transfer or superpose the heralded mechanical states into or with the optical domain at $\omega_{j+2}$, where they can be isolated using band-pass filters. 
Physically, these stop-bands can be engineered in a multitude of ways, including Bragg gratings \cite{lee_suppression_2003, merklein_enhancing_2015} or structural changes to the resonator system \cite{puckett_higher_2019, wang_taming_2024}.

%\section{Outlook}
In summary, we show how the quantum dynamics of intra-modal spontaneous forward Brillouin scattering can be paired with single photon detection to herald phonon Fock states of an arbitrary and selective number. Exploiting the entanglement generated by the system when stimulated with a single photon or a faint laser, single-photon detection can collapse the state into a phonon Fock state, with the phonon number being determined by the relative frequency shift of the detected photon with respect to the optical input. We discuss a potential method for quantum state tomography of the heralded phonon, through engineered optical mode suppression, which enables optomechanical $\pi$- and $\pi/2$- pulses.

Mechanical decoherence is ignored, as this analysis only considers times far shorter than reported phonon coherence times, and although optical losses can be significant, a lost photon will only lower the probability of detection, leaving the heralded multi-phonon Fock state unaffected.
This result, in conjunction with long phonon quantum coherence times, the compatibility of these systems with telecommunication wavelengths of light, the large effective mass of the phonon modes, and access to a ladder of resonant frequencies of light, has the potential to enable new quantum devices, new approaches to state synthesis, and open new windows on fundamental physics questions.

{\it Acknowledgments}---
This work was supported by NSF Award No. 2145724.

%\bibliography{references}

\input{Main.bbl}

\end{document}

%% file: Main.bbl

%% file: Main.bbl
\begin{thebibliography}{45}%
\makeatletter
\providecommand \@ifxundefined [1]{%
 \@ifx{#1\undefined}
}%
\providecommand \@ifnum [1]{%
 \ifnum #1\expandafter \@firstoftwo
 \else \expandafter \@secondoftwo
 \fi
}%
\providecommand \@ifx [1]{%
 \ifx #1\expandafter \@firstoftwo
 \else \expandafter \@secondoftwo
 \fi
}%
\providecommand \natexlab [1]{#1}%
\providecommand \enquote  [1]{``#1''}%
\providecommand \bibnamefont  [1]{#1}%
\providecommand \bibfnamefont [1]{#1}%
\providecommand \citenamefont [1]{#1}%
\providecommand \href@noop [0]{\@secondoftwo}%
\providecommand \href [0]{\begingroup \@sanitize@url \@href}%
\providecommand \@href[1]{\@@startlink{#1}\@@href}%
\providecommand \@@href[1]{\endgroup#1\@@endlink}%
\providecommand \@sanitize@url [0]{\catcode `\\12\catcode `\$12\catcode `\&12\catcode `\#12\catcode `\^12\catcode `\_12\catcode `\%12\relax}%
\providecommand \@@startlink[1]{}%
\providecommand \@@endlink[0]{}%
\providecommand \url  [0]{\begingroup\@sanitize@url \@url }%
\providecommand \@url [1]{\endgroup\@href {#1}{\urlprefix }}%
\providecommand \urlprefix  [0]{URL }%
\providecommand \Eprint [0]{\href }%
\providecommand \doibase [0]{http://dx.doi.org/}%
\providecommand \selectlanguage [0]{\@gobble}%
\providecommand \bibinfo  [0]{\@secondoftwo}%
\providecommand \bibfield  [0]{\@secondoftwo}%
\providecommand \translation [1]{[#1]}%
\providecommand \BibitemOpen [0]{}%
\providecommand \bibitemStop [0]{}%
\providecommand \bibitemNoStop [0]{.\EOS\space}%
\providecommand \EOS [0]{\spacefactor3000\relax}%
\providecommand \BibitemShut  [1]{\csname bibitem#1\endcsname}%
\let\auto@bib@innerbib\@empty
%</preamble>
\bibitem [{\citenamefont {Couteau}(2018)}]{couteau_spontaneous_2018}%
  \BibitemOpen
  \bibfield  {author} {\bibinfo {author} {\bibfnamefont {C.}~\bibnamefont {Couteau}},\ }\href {\doibase 10.1080/00107514.2018.1488463} {\bibfield  {journal} {\bibinfo  {journal} {Contemporary Physics}\ }\textbf {\bibinfo {volume} {59}},\ \bibinfo {pages} {291} (\bibinfo {year} {2018})},\ \bibinfo {note} {arXiv:1809.00127 [physics, physics:quant-ph]}\BibitemShut {NoStop}%
\bibitem [{\citenamefont {Caspani}\ \emph {et~al.}(2017)\citenamefont {Caspani}, \citenamefont {Xiong}, \citenamefont {Eggleton}, \citenamefont {Bajoni}, \citenamefont {Liscidini}, \citenamefont {Galli}, \citenamefont {Morandotti},\ and\ \citenamefont {Moss}}]{caspani_integrated_2017}%
  \BibitemOpen
  \bibfield  {author} {\bibinfo {author} {\bibfnamefont {L.}~\bibnamefont {Caspani}}, \bibinfo {author} {\bibfnamefont {C.}~\bibnamefont {Xiong}}, \bibinfo {author} {\bibfnamefont {B.~J.}\ \bibnamefont {Eggleton}}, \bibinfo {author} {\bibfnamefont {D.}~\bibnamefont {Bajoni}}, \bibinfo {author} {\bibfnamefont {M.}~\bibnamefont {Liscidini}}, \bibinfo {author} {\bibfnamefont {M.}~\bibnamefont {Galli}}, \bibinfo {author} {\bibfnamefont {R.}~\bibnamefont {Morandotti}}, \ and\ \bibinfo {author} {\bibfnamefont {D.~J.}\ \bibnamefont {Moss}},\ }\href {\doibase 10.1038/lsa.2017.100} {\bibfield  {journal} {\bibinfo  {journal} {Light: Science \& Applications}\ }\textbf {\bibinfo {volume} {6}},\ \bibinfo {pages} {e17100} (\bibinfo {year} {2017})},\ \bibinfo {note} {number: 11 Publisher: Nature Publishing Group}\BibitemShut {NoStop}%
\bibitem [{\citenamefont {Zavatta}\ \emph {et~al.}(2004)\citenamefont {Zavatta}, \citenamefont {Viciani},\ and\ \citenamefont {Bellini}}]{zavatta_quantum--classical_2004}%
  \BibitemOpen
  \bibfield  {author} {\bibinfo {author} {\bibfnamefont {A.}~\bibnamefont {Zavatta}}, \bibinfo {author} {\bibfnamefont {S.}~\bibnamefont {Viciani}}, \ and\ \bibinfo {author} {\bibfnamefont {M.}~\bibnamefont {Bellini}},\ }\href {\doibase 10.1126/science.1103190} {\bibfield  {journal} {\bibinfo  {journal} {Science (New York, N.Y.)}\ }\textbf {\bibinfo {volume} {306}},\ \bibinfo {pages} {660} (\bibinfo {year} {2004})}\BibitemShut {NoStop}%
\bibitem [{\citenamefont {Hong}\ \emph {et~al.}(2017)\citenamefont {Hong}, \citenamefont {Riedinger}, \citenamefont {Marinković}, \citenamefont {Wallucks}, \citenamefont {Hofer}, \citenamefont {Norte}, \citenamefont {Aspelmeyer},\ and\ \citenamefont {Gröblacher}}]{hong_hanbury_2017}%
  \BibitemOpen
  \bibfield  {author} {\bibinfo {author} {\bibfnamefont {S.}~\bibnamefont {Hong}}, \bibinfo {author} {\bibfnamefont {R.}~\bibnamefont {Riedinger}}, \bibinfo {author} {\bibfnamefont {I.}~\bibnamefont {Marinković}}, \bibinfo {author} {\bibfnamefont {A.}~\bibnamefont {Wallucks}}, \bibinfo {author} {\bibfnamefont {S.~G.}\ \bibnamefont {Hofer}}, \bibinfo {author} {\bibfnamefont {R.~A.}\ \bibnamefont {Norte}}, \bibinfo {author} {\bibfnamefont {M.}~\bibnamefont {Aspelmeyer}}, \ and\ \bibinfo {author} {\bibfnamefont {S.}~\bibnamefont {Gröblacher}},\ }\href {\doibase 10.1126/science.aan7939} {\bibfield  {journal} {\bibinfo  {journal} {Science}\ }\textbf {\bibinfo {volume} {358}},\ \bibinfo {pages} {203} (\bibinfo {year} {2017})}\BibitemShut {NoStop}%
\bibitem [{\citenamefont {Galland}\ \emph {et~al.}(2014)\citenamefont {Galland}, \citenamefont {Sangouard}, \citenamefont {Piro}, \citenamefont {Gisin},\ and\ \citenamefont {Kippenberg}}]{galland_heralded_2014}%
  \BibitemOpen
  \bibfield  {author} {\bibinfo {author} {\bibfnamefont {C.}~\bibnamefont {Galland}}, \bibinfo {author} {\bibfnamefont {N.}~\bibnamefont {Sangouard}}, \bibinfo {author} {\bibfnamefont {N.}~\bibnamefont {Piro}}, \bibinfo {author} {\bibfnamefont {N.}~\bibnamefont {Gisin}}, \ and\ \bibinfo {author} {\bibfnamefont {T.~J.}\ \bibnamefont {Kippenberg}},\ }\href {\doibase 10.1103/PhysRevLett.112.143602} {\bibfield  {journal} {\bibinfo  {journal} {Physical Review Letters}\ }\textbf {\bibinfo {volume} {112}},\ \bibinfo {pages} {143602} (\bibinfo {year} {2014})}\BibitemShut {NoStop}%
\bibitem [{\citenamefont {Goryachev}\ \emph {et~al.}(2012)\citenamefont {Goryachev}, \citenamefont {Creedon}, \citenamefont {Ivanov}, \citenamefont {Galliou}, \citenamefont {Bourquin},\ and\ \citenamefont {Tobar}}]{goryachev_extremely_2012}%
  \BibitemOpen
  \bibfield  {author} {\bibinfo {author} {\bibfnamefont {M.}~\bibnamefont {Goryachev}}, \bibinfo {author} {\bibfnamefont {D.~L.}\ \bibnamefont {Creedon}}, \bibinfo {author} {\bibfnamefont {E.~N.}\ \bibnamefont {Ivanov}}, \bibinfo {author} {\bibfnamefont {S.}~\bibnamefont {Galliou}}, \bibinfo {author} {\bibfnamefont {R.}~\bibnamefont {Bourquin}}, \ and\ \bibinfo {author} {\bibfnamefont {M.~E.}\ \bibnamefont {Tobar}},\ }\href {\doibase 10.1063/1.4729292} {\bibfield  {journal} {\bibinfo  {journal} {Applied Physics Letters}\ }\textbf {\bibinfo {volume} {100}},\ \bibinfo {pages} {243504} (\bibinfo {year} {2012})},\ \bibinfo {note} {publisher: American Institute of Physics}\BibitemShut {NoStop}%
\bibitem [{\citenamefont {Goryachev}\ \emph {et~al.}(2013)\citenamefont {Goryachev}, \citenamefont {Creedon}, \citenamefont {Galliou},\ and\ \citenamefont {Tobar}}]{goryachev_observation_2013}%
  \BibitemOpen
  \bibfield  {author} {\bibinfo {author} {\bibfnamefont {M.}~\bibnamefont {Goryachev}}, \bibinfo {author} {\bibfnamefont {D.~L.}\ \bibnamefont {Creedon}}, \bibinfo {author} {\bibfnamefont {S.}~\bibnamefont {Galliou}}, \ and\ \bibinfo {author} {\bibfnamefont {M.~E.}\ \bibnamefont {Tobar}},\ }\href {\doibase 10.1103/PhysRevLett.111.085502} {\bibfield  {journal} {\bibinfo  {journal} {Physical Review Letters}\ }\textbf {\bibinfo {volume} {111}},\ \bibinfo {pages} {085502} (\bibinfo {year} {2013})},\ \bibinfo {note} {publisher: American Physical Society}\BibitemShut {NoStop}%
\bibitem [{\citenamefont {MacCabe}\ \emph {et~al.}(2020)\citenamefont {MacCabe}, \citenamefont {Ren}, \citenamefont {Luo}, \citenamefont {Cohen}, \citenamefont {Zhou}, \citenamefont {Sipahigil}, \citenamefont {Mirhosseini},\ and\ \citenamefont {Painter}}]{maccabe_nano-acoustic_2020}%
  \BibitemOpen
  \bibfield  {author} {\bibinfo {author} {\bibfnamefont {G.~S.}\ \bibnamefont {MacCabe}}, \bibinfo {author} {\bibfnamefont {H.}~\bibnamefont {Ren}}, \bibinfo {author} {\bibfnamefont {J.}~\bibnamefont {Luo}}, \bibinfo {author} {\bibfnamefont {J.~D.}\ \bibnamefont {Cohen}}, \bibinfo {author} {\bibfnamefont {H.}~\bibnamefont {Zhou}}, \bibinfo {author} {\bibfnamefont {A.}~\bibnamefont {Sipahigil}}, \bibinfo {author} {\bibfnamefont {M.}~\bibnamefont {Mirhosseini}}, \ and\ \bibinfo {author} {\bibfnamefont {O.}~\bibnamefont {Painter}},\ }\href {\doibase 10.1126/science.abc7312} {\bibfield  {journal} {\bibinfo  {journal} {Science}\ }\textbf {\bibinfo {volume} {370}},\ \bibinfo {pages} {840} (\bibinfo {year} {2020})}\BibitemShut {NoStop}%
\bibitem [{\citenamefont {{K. J. Satzinger}}\ \emph {et~al.}(2018)\citenamefont {{K. J. Satzinger}}, \citenamefont {Zhong}, \citenamefont {Chang}, \citenamefont {Peairs}, \citenamefont {Bienfait}, \citenamefont {Chou}, \citenamefont {Cleland}, \citenamefont {Conner}, \citenamefont {Dumur}, \citenamefont {Grebel}, \citenamefont {Gutierrez}, \citenamefont {November}, \citenamefont {Povey}, \citenamefont {Whiteley}, \citenamefont {Awschalom}, \citenamefont {Schuster},\ and\ \citenamefont {Cleland}}]{k_j_satzinger_quantum_2018}%
  \BibitemOpen
  \bibfield  {author} {\bibinfo {author} {\bibnamefont {{K. J. Satzinger}}}, \bibinfo {author} {\bibfnamefont {Y.~P.}\ \bibnamefont {Zhong}}, \bibinfo {author} {\bibfnamefont {H.-S.}\ \bibnamefont {Chang}}, \bibinfo {author} {\bibfnamefont {G.~A.}\ \bibnamefont {Peairs}}, \bibinfo {author} {\bibfnamefont {A.}~\bibnamefont {Bienfait}}, \bibinfo {author} {\bibfnamefont {M.-H.}\ \bibnamefont {Chou}}, \bibinfo {author} {\bibfnamefont {A.~Y.}\ \bibnamefont {Cleland}}, \bibinfo {author} {\bibfnamefont {C.~R.}\ \bibnamefont {Conner}}, \bibinfo {author} {\bibfnamefont {Ã.}~\bibnamefont {Dumur}}, \bibinfo {author} {\bibfnamefont {J.}~\bibnamefont {Grebel}}, \bibinfo {author} {\bibfnamefont {I.}~\bibnamefont {Gutierrez}}, \bibinfo {author} {\bibfnamefont {B.~H.}\ \bibnamefont {November}}, \bibinfo {author} {\bibfnamefont {R.~G.}\ \bibnamefont {Povey}}, \bibinfo {author} {\bibfnamefont {S.~J.}\ \bibnamefont {Whiteley}}, \bibinfo {author} {\bibfnamefont {D.~D.}\ \bibnamefont {Awschalom}}, \bibinfo {author} {\bibfnamefont
  {D.~I.}\ \bibnamefont {Schuster}}, \ and\ \bibinfo {author} {\bibfnamefont {A.~N.}\ \bibnamefont {Cleland}},\ }\href {\doibase 10.1038/s41586-018-0719-5} {\bibfield  {journal} {\bibinfo  {journal} {Nature}\ }\textbf {\bibinfo {volume} {563}},\ \bibinfo {pages} {661} (\bibinfo {year} {2018})},\ \bibinfo {note} {number: 7733 Publisher: Nature Publishing Group}\BibitemShut {NoStop}%
\bibitem [{\citenamefont {Chu}\ \emph {et~al.}(2018)\citenamefont {Chu}, \citenamefont {Kharel}, \citenamefont {Yoon}, \citenamefont {Frunzio}, \citenamefont {Rakich},\ and\ \citenamefont {Schoelkopf}}]{chu_creation_2018}%
  \BibitemOpen
  \bibfield  {author} {\bibinfo {author} {\bibfnamefont {Y.}~\bibnamefont {Chu}}, \bibinfo {author} {\bibfnamefont {P.}~\bibnamefont {Kharel}}, \bibinfo {author} {\bibfnamefont {T.}~\bibnamefont {Yoon}}, \bibinfo {author} {\bibfnamefont {L.}~\bibnamefont {Frunzio}}, \bibinfo {author} {\bibfnamefont {P.~T.}\ \bibnamefont {Rakich}}, \ and\ \bibinfo {author} {\bibfnamefont {R.~J.}\ \bibnamefont {Schoelkopf}},\ }\href {\doibase 10.1038/s41586-018-0717-7} {\bibfield  {journal} {\bibinfo  {journal} {Nature}\ }\textbf {\bibinfo {volume} {563}},\ \bibinfo {pages} {666} (\bibinfo {year} {2018})},\ \bibinfo {note} {publisher: Nature Publishing Group}\BibitemShut {NoStop}%
\bibitem [{\citenamefont {Stannigel}\ \emph {et~al.}(2010)\citenamefont {Stannigel}, \citenamefont {Rabl}, \citenamefont {Sørensen}, \citenamefont {Zoller},\ and\ \citenamefont {Lukin}}]{stannigel_optomechanical_2010}%
  \BibitemOpen
  \bibfield  {author} {\bibinfo {author} {\bibfnamefont {K.}~\bibnamefont {Stannigel}}, \bibinfo {author} {\bibfnamefont {P.}~\bibnamefont {Rabl}}, \bibinfo {author} {\bibfnamefont {A.~S.}\ \bibnamefont {Sørensen}}, \bibinfo {author} {\bibfnamefont {P.}~\bibnamefont {Zoller}}, \ and\ \bibinfo {author} {\bibfnamefont {M.~D.}\ \bibnamefont {Lukin}},\ }\href {\doibase 10.1103/PhysRevLett.105.220501} {\bibfield  {journal} {\bibinfo  {journal} {Physical Review Letters}\ }\textbf {\bibinfo {volume} {105}},\ \bibinfo {pages} {220501} (\bibinfo {year} {2010})},\ \bibinfo {note} {publisher: American Physical Society}\BibitemShut {NoStop}%
\bibitem [{\citenamefont {Jiang}\ \emph {et~al.}(2020)\citenamefont {Jiang}, \citenamefont {Sarabalis}, \citenamefont {Dahmani}, \citenamefont {Patel}, \citenamefont {Mayor}, \citenamefont {McKenna}, \citenamefont {Van~Laer},\ and\ \citenamefont {Safavi-Naeini}}]{jiang_efficient_2020}%
  \BibitemOpen
  \bibfield  {author} {\bibinfo {author} {\bibfnamefont {W.}~\bibnamefont {Jiang}}, \bibinfo {author} {\bibfnamefont {C.~J.}\ \bibnamefont {Sarabalis}}, \bibinfo {author} {\bibfnamefont {Y.~D.}\ \bibnamefont {Dahmani}}, \bibinfo {author} {\bibfnamefont {R.~N.}\ \bibnamefont {Patel}}, \bibinfo {author} {\bibfnamefont {F.~M.}\ \bibnamefont {Mayor}}, \bibinfo {author} {\bibfnamefont {T.~P.}\ \bibnamefont {McKenna}}, \bibinfo {author} {\bibfnamefont {R.}~\bibnamefont {Van~Laer}}, \ and\ \bibinfo {author} {\bibfnamefont {A.~H.}\ \bibnamefont {Safavi-Naeini}},\ }\href {\doibase 10.1038/s41467-020-14863-3} {\bibfield  {journal} {\bibinfo  {journal} {Nature Communications}\ }\textbf {\bibinfo {volume} {11}},\ \bibinfo {pages} {1166} (\bibinfo {year} {2020})},\ \bibinfo {note} {publisher: Nature Publishing Group}\BibitemShut {NoStop}%
\bibitem [{\citenamefont {Pechal}\ \emph {et~al.}(2018)\citenamefont {Pechal}, \citenamefont {Arrangoiz-Arriola},\ and\ \citenamefont {Safavi-Naeini}}]{pechal_superconducting_2018}%
  \BibitemOpen
  \bibfield  {author} {\bibinfo {author} {\bibfnamefont {M.}~\bibnamefont {Pechal}}, \bibinfo {author} {\bibfnamefont {P.}~\bibnamefont {Arrangoiz-Arriola}}, \ and\ \bibinfo {author} {\bibfnamefont {A.~H.}\ \bibnamefont {Safavi-Naeini}},\ }\href {\doibase 10.1088/2058-9565/aadc6c} {\bibfield  {journal} {\bibinfo  {journal} {Quantum Science and Technology}\ }\textbf {\bibinfo {volume} {4}},\ \bibinfo {pages} {015006} (\bibinfo {year} {2018})},\ \bibinfo {note} {publisher: IOP Publishing}\BibitemShut {NoStop}%
\bibitem [{\citenamefont {Mason}\ \emph {et~al.}(2019)\citenamefont {Mason}, \citenamefont {Chen}, \citenamefont {Rossi}, \citenamefont {Tsaturyan},\ and\ \citenamefont {Schliesser}}]{mason_continuous_2019}%
  \BibitemOpen
  \bibfield  {author} {\bibinfo {author} {\bibfnamefont {D.}~\bibnamefont {Mason}}, \bibinfo {author} {\bibfnamefont {J.}~\bibnamefont {Chen}}, \bibinfo {author} {\bibfnamefont {M.}~\bibnamefont {Rossi}}, \bibinfo {author} {\bibfnamefont {Y.}~\bibnamefont {Tsaturyan}}, \ and\ \bibinfo {author} {\bibfnamefont {A.}~\bibnamefont {Schliesser}},\ }\href {\doibase 10.1038/s41567-019-0533-5} {\bibfield  {journal} {\bibinfo  {journal} {Nature Physics}\ }\textbf {\bibinfo {volume} {15}},\ \bibinfo {pages} {745} (\bibinfo {year} {2019})},\ \bibinfo {note} {publisher: Nature Publishing Group}\BibitemShut {NoStop}%
\bibitem [{\citenamefont {Marshall}\ \emph {et~al.}(2003)\citenamefont {Marshall}, \citenamefont {Simon}, \citenamefont {Penrose},\ and\ \citenamefont {Bouwmeester}}]{marshall_towards_2003}%
  \BibitemOpen
  \bibfield  {author} {\bibinfo {author} {\bibfnamefont {W.}~\bibnamefont {Marshall}}, \bibinfo {author} {\bibfnamefont {C.}~\bibnamefont {Simon}}, \bibinfo {author} {\bibfnamefont {R.}~\bibnamefont {Penrose}}, \ and\ \bibinfo {author} {\bibfnamefont {D.}~\bibnamefont {Bouwmeester}},\ }\href {\doibase 10.1103/PhysRevLett.91.130401} {\bibfield  {journal} {\bibinfo  {journal} {Physical Review Letters}\ }\textbf {\bibinfo {volume} {91}},\ \bibinfo {pages} {130401} (\bibinfo {year} {2003})},\ \bibinfo {note} {publisher: American Physical Society}\BibitemShut {NoStop}%
\bibitem [{\citenamefont {Diósi}(1989)}]{diosi_models_1989}%
  \BibitemOpen
  \bibfield  {author} {\bibinfo {author} {\bibfnamefont {L.}~\bibnamefont {Diósi}},\ }\href {\doibase 10.1103/PhysRevA.40.1165} {\bibfield  {journal} {\bibinfo  {journal} {Physical Review A}\ }\textbf {\bibinfo {volume} {40}},\ \bibinfo {pages} {1165} (\bibinfo {year} {1989})},\ \bibinfo {note} {publisher: American Physical Society}\BibitemShut {NoStop}%
\bibitem [{\citenamefont {Penrose}(2014)}]{penrose_gravitization_2014}%
  \BibitemOpen
  \bibfield  {author} {\bibinfo {author} {\bibfnamefont {R.}~\bibnamefont {Penrose}},\ }\href {\doibase 10.1007/s10701-013-9770-0} {\bibfield  {journal} {\bibinfo  {journal} {Foundations of Physics}\ }\textbf {\bibinfo {volume} {44}},\ \bibinfo {pages} {557} (\bibinfo {year} {2014})}\BibitemShut {NoStop}%
\bibitem [{\citenamefont {Ghirardi}\ \emph {et~al.}(1990)\citenamefont {Ghirardi}, \citenamefont {Pearle},\ and\ \citenamefont {Rimini}}]{ghirardi_markov_1990}%
  \BibitemOpen
  \bibfield  {author} {\bibinfo {author} {\bibfnamefont {G.~C.}\ \bibnamefont {Ghirardi}}, \bibinfo {author} {\bibfnamefont {P.}~\bibnamefont {Pearle}}, \ and\ \bibinfo {author} {\bibfnamefont {A.}~\bibnamefont {Rimini}},\ }\href {\doibase 10.1103/PhysRevA.42.78} {\bibfield  {journal} {\bibinfo  {journal} {Physical Review A}\ }\textbf {\bibinfo {volume} {42}},\ \bibinfo {pages} {78} (\bibinfo {year} {1990})},\ \bibinfo {note} {publisher: American Physical Society}\BibitemShut {NoStop}%
\bibitem [{\citenamefont {Chu}\ \emph {et~al.}(2017)\citenamefont {Chu}, \citenamefont {Kharel}, \citenamefont {Renninger}, \citenamefont {Burkhart}, \citenamefont {Frunzio}, \citenamefont {Rakich},\ and\ \citenamefont {Schoelkopf}}]{chu_quantum_2017}%
  \BibitemOpen
  \bibfield  {author} {\bibinfo {author} {\bibfnamefont {Y.}~\bibnamefont {Chu}}, \bibinfo {author} {\bibfnamefont {P.}~\bibnamefont {Kharel}}, \bibinfo {author} {\bibfnamefont {W.~H.}\ \bibnamefont {Renninger}}, \bibinfo {author} {\bibfnamefont {L.~D.}\ \bibnamefont {Burkhart}}, \bibinfo {author} {\bibfnamefont {L.}~\bibnamefont {Frunzio}}, \bibinfo {author} {\bibfnamefont {P.~T.}\ \bibnamefont {Rakich}}, \ and\ \bibinfo {author} {\bibfnamefont {R.~J.}\ \bibnamefont {Schoelkopf}},\ }\href {\doibase 10.1126/science.aao1511} {\bibfield  {journal} {\bibinfo  {journal} {Science}\ }\textbf {\bibinfo {volume} {358}},\ \bibinfo {pages} {199} (\bibinfo {year} {2017})},\ \bibinfo {note} {publisher: American Association for the Advancement of Science}\BibitemShut {NoStop}%
\bibitem [{\citenamefont {Behunin}\ and\ \citenamefont {Rakich}(2023)}]{behunin_harnessing_2023}%
  \BibitemOpen
  \bibfield  {author} {\bibinfo {author} {\bibfnamefont {R.~O.}\ \bibnamefont {Behunin}}\ and\ \bibinfo {author} {\bibfnamefont {P.~T.}\ \bibnamefont {Rakich}},\ }\href {\doibase 10.1103/PhysRevA.107.023511} {\bibfield  {journal} {\bibinfo  {journal} {Physical Review A}\ }\textbf {\bibinfo {volume} {107}},\ \bibinfo {pages} {023511} (\bibinfo {year} {2023})},\ \bibinfo {note} {publisher: American Physical Society}\BibitemShut {NoStop}%
\bibitem [{\citenamefont {Shelby}\ \emph {et~al.}(1985)\citenamefont {Shelby}, \citenamefont {Levenson},\ and\ \citenamefont {Bayer}}]{shelby_resolved_1985}%
  \BibitemOpen
  \bibfield  {author} {\bibinfo {author} {\bibfnamefont {R.~M.}\ \bibnamefont {Shelby}}, \bibinfo {author} {\bibfnamefont {M.~D.}\ \bibnamefont {Levenson}}, \ and\ \bibinfo {author} {\bibfnamefont {P.~W.}\ \bibnamefont {Bayer}},\ }\href {\doibase 10.1103/PhysRevLett.54.939} {\bibfield  {journal} {\bibinfo  {journal} {Physical Review Letters}\ }\textbf {\bibinfo {volume} {54}},\ \bibinfo {pages} {939} (\bibinfo {year} {1985})},\ \bibinfo {note} {publisher: American Physical Society}\BibitemShut {NoStop}%
\bibitem [{\citenamefont {Raymer}\ and\ \citenamefont {Mostowski}(1981)}]{raymer_stimulated_1981}%
  \BibitemOpen
  \bibfield  {author} {\bibinfo {author} {\bibfnamefont {M.~G.}\ \bibnamefont {Raymer}}\ and\ \bibinfo {author} {\bibfnamefont {J.}~\bibnamefont {Mostowski}},\ }\href {\doibase 10.1103/PhysRevA.24.1980} {\bibfield  {journal} {\bibinfo  {journal} {Physical Review A}\ }\textbf {\bibinfo {volume} {24}},\ \bibinfo {pages} {1980} (\bibinfo {year} {1981})},\ \bibinfo {note} {publisher: American Physical Society}\BibitemShut {NoStop}%
\bibitem [{\citenamefont {Kharel}\ \emph {et~al.}(2016)\citenamefont {Kharel}, \citenamefont {Behunin}, \citenamefont {Renninger},\ and\ \citenamefont {Rakich}}]{kharel_noise_2016}%
  \BibitemOpen
  \bibfield  {author} {\bibinfo {author} {\bibfnamefont {P.}~\bibnamefont {Kharel}}, \bibinfo {author} {\bibfnamefont {R.~O.}\ \bibnamefont {Behunin}}, \bibinfo {author} {\bibfnamefont {W.~H.}\ \bibnamefont {Renninger}}, \ and\ \bibinfo {author} {\bibfnamefont {P.~T.}\ \bibnamefont {Rakich}},\ }\href {\doibase 10.1103/PhysRevA.93.063806} {\bibfield  {journal} {\bibinfo  {journal} {Physical Review A}\ }\textbf {\bibinfo {volume} {93}},\ \bibinfo {pages} {063806} (\bibinfo {year} {2016})},\ \bibinfo {note} {publisher: American Physical Society}\BibitemShut {NoStop}%
\bibitem [{\citenamefont {Boyd}(2008)}]{boyd_chapter_2008}%
  \BibitemOpen
  \bibfield  {author} {\bibinfo {author} {\bibfnamefont {R.~W.}\ \bibnamefont {Boyd}},\ }in\ \href {\doibase 10.1016/B978-0-12-369470-6.00009-5} {\emph {\bibinfo {booktitle} {Nonlinear {Optics} ({Third} {Edition})}}},\ \bibinfo {editor} {edited by\ \bibinfo {editor} {\bibfnamefont {R.~W.}\ \bibnamefont {Boyd}}}\ (\bibinfo  {publisher} {Academic Press},\ \bibinfo {address} {Burlington},\ \bibinfo {year} {2008})\ pp.\ \bibinfo {pages} {429--471}\BibitemShut {NoStop}%
\bibitem [{\citenamefont {Shanavas}\ \emph {et~al.}(2022)\citenamefont {Shanavas}, \citenamefont {Grayson}, \citenamefont {Zohrabi}, \citenamefont {Park},\ and\ \citenamefont {Gopinath}}]{shanavas_cascaded_2022}%
  \BibitemOpen
  \bibfield  {author} {\bibinfo {author} {\bibfnamefont {T.}~\bibnamefont {Shanavas}}, \bibinfo {author} {\bibfnamefont {M.~B.}\ \bibnamefont {Grayson}}, \bibinfo {author} {\bibfnamefont {M.}~\bibnamefont {Zohrabi}}, \bibinfo {author} {\bibfnamefont {W.}~\bibnamefont {Park}}, \ and\ \bibinfo {author} {\bibfnamefont {J.~T.}\ \bibnamefont {Gopinath}},\ }in\ \href {\doibase 10.1364/CLEO_SI.2022.STh5F.7} {{\selectlanguage {EN}\emph {\bibinfo {booktitle} {Conference on {Lasers} and {Electro}-{Optics} (2022), paper {STh5F}.7}}}}\ (\bibinfo  {publisher} {Optica Publishing Group},\ \bibinfo {year} {2022})\ p.\ \bibinfo {pages} {STh5F.7}\BibitemShut {NoStop}%
\bibitem [{\citenamefont {Rakich}\ \emph {et~al.}(2012)\citenamefont {Rakich}, \citenamefont {Reinke}, \citenamefont {Camacho}, \citenamefont {Davids},\ and\ \citenamefont {Wang}}]{rakich_giant_2012}%
  \BibitemOpen
  \bibfield  {author} {\bibinfo {author} {\bibfnamefont {P.~T.}\ \bibnamefont {Rakich}}, \bibinfo {author} {\bibfnamefont {C.}~\bibnamefont {Reinke}}, \bibinfo {author} {\bibfnamefont {R.}~\bibnamefont {Camacho}}, \bibinfo {author} {\bibfnamefont {P.}~\bibnamefont {Davids}}, \ and\ \bibinfo {author} {\bibfnamefont {Z.}~\bibnamefont {Wang}},\ }\href {\doibase 10.1103/PhysRevX.2.011008} {\bibfield  {journal} {\bibinfo  {journal} {Physical Review X}\ }\textbf {\bibinfo {volume} {2}},\ \bibinfo {pages} {011008} (\bibinfo {year} {2012})}\BibitemShut {NoStop}%
\bibitem [{\citenamefont {Eggleton}\ \emph {et~al.}(2019)\citenamefont {Eggleton}, \citenamefont {Poulton}, \citenamefont {Rakich}, \citenamefont {Steel},\ and\ \citenamefont {Bahl}}]{eggleton_brillouin_2019}%
  \BibitemOpen
  \bibfield  {author} {\bibinfo {author} {\bibfnamefont {B.~J.}\ \bibnamefont {Eggleton}}, \bibinfo {author} {\bibfnamefont {C.~G.}\ \bibnamefont {Poulton}}, \bibinfo {author} {\bibfnamefont {P.~T.}\ \bibnamefont {Rakich}}, \bibinfo {author} {\bibfnamefont {M.~J.}\ \bibnamefont {Steel}}, \ and\ \bibinfo {author} {\bibfnamefont {G.}~\bibnamefont {Bahl}},\ }\href {\doibase 10.1038/s41566-019-0498-z} {\bibfield  {journal} {\bibinfo  {journal} {Nature Photonics}\ }\textbf {\bibinfo {volume} {13}},\ \bibinfo {pages} {664} (\bibinfo {year} {2019})},\ \bibinfo {note} {publisher: Nature Publishing Group}\BibitemShut {NoStop}%
\bibitem [{\citenamefont {Kang}\ \emph {et~al.}(2009)\citenamefont {Kang}, \citenamefont {Nazarkin}, \citenamefont {Brenn},\ and\ \citenamefont {Russell}}]{kang_tightly_2009}%
  \BibitemOpen
  \bibfield  {author} {\bibinfo {author} {\bibfnamefont {M.~S.}\ \bibnamefont {Kang}}, \bibinfo {author} {\bibfnamefont {A.}~\bibnamefont {Nazarkin}}, \bibinfo {author} {\bibfnamefont {A.}~\bibnamefont {Brenn}}, \ and\ \bibinfo {author} {\bibfnamefont {P.~S.~J.}\ \bibnamefont {Russell}},\ }\href {\doibase 10.1038/nphys1217} {\bibfield  {journal} {\bibinfo  {journal} {Nature Physics}\ }\textbf {\bibinfo {volume} {5}},\ \bibinfo {pages} {276} (\bibinfo {year} {2009})},\ \bibinfo {note} {number: 4 Publisher: Nature Publishing Group}\BibitemShut {NoStop}%
\bibitem [{\citenamefont {Kang}\ \emph {et~al.}(2010)\citenamefont {Kang}, \citenamefont {Brenn},\ and\ \citenamefont {St.J.~Russell}}]{kang_all-optical_2010}%
  \BibitemOpen
  \bibfield  {author} {\bibinfo {author} {\bibfnamefont {M.~S.}\ \bibnamefont {Kang}}, \bibinfo {author} {\bibfnamefont {A.}~\bibnamefont {Brenn}}, \ and\ \bibinfo {author} {\bibfnamefont {P.}~\bibnamefont {St.J.~Russell}},\ }\href {\doibase 10.1103/PhysRevLett.105.153901} {\bibfield  {journal} {\bibinfo  {journal} {Physical Review Letters}\ }\textbf {\bibinfo {volume} {105}},\ \bibinfo {pages} {153901} (\bibinfo {year} {2010})}\BibitemShut {NoStop}%
\bibitem [{\citenamefont {Renninger}\ \emph {et~al.}(2016{\natexlab{a}})\citenamefont {Renninger}, \citenamefont {Behunin},\ and\ \citenamefont {Rakich}}]{renninger_guided-wave_2016}%
  \BibitemOpen
  \bibfield  {author} {\bibinfo {author} {\bibfnamefont {W.~H.}\ \bibnamefont {Renninger}}, \bibinfo {author} {\bibfnamefont {R.~O.}\ \bibnamefont {Behunin}}, \ and\ \bibinfo {author} {\bibfnamefont {P.~T.}\ \bibnamefont {Rakich}},\ }\href {\doibase 10.1364/OPTICA.3.001316} {\bibfield  {journal} {\bibinfo  {journal} {Optica}\ }\textbf {\bibinfo {volume} {3}},\ \bibinfo {pages} {1316} (\bibinfo {year} {2016}{\natexlab{a}})},\ \bibinfo {note} {publisher: Optica Publishing Group}\BibitemShut {NoStop}%
\bibitem [{\citenamefont {Renninger}\ \emph {et~al.}(2016{\natexlab{b}})\citenamefont {Renninger}, \citenamefont {Shin}, \citenamefont {Behunin}, \citenamefont {Kharel}, \citenamefont {Kittlaus},\ and\ \citenamefont {Rakich}}]{renninger_forward_2016}%
  \BibitemOpen
  \bibfield  {author} {\bibinfo {author} {\bibfnamefont {W.~H.}\ \bibnamefont {Renninger}}, \bibinfo {author} {\bibfnamefont {H.}~\bibnamefont {Shin}}, \bibinfo {author} {\bibfnamefont {R.~O.}\ \bibnamefont {Behunin}}, \bibinfo {author} {\bibfnamefont {P.}~\bibnamefont {Kharel}}, \bibinfo {author} {\bibfnamefont {E.~A.}\ \bibnamefont {Kittlaus}}, \ and\ \bibinfo {author} {\bibfnamefont {P.~T.}\ \bibnamefont {Rakich}},\ }\href {\doibase 10.1088/1367-2630/18/2/025008} {\bibfield  {journal} {\bibinfo  {journal} {New Journal of Physics}\ }\textbf {\bibinfo {volume} {18}},\ \bibinfo {pages} {025008} (\bibinfo {year} {2016}{\natexlab{b}})},\ \bibinfo {note} {publisher: IOP Publishing}\BibitemShut {NoStop}%
\bibitem [{\citenamefont {Behunin}\ \emph {et~al.}(2019)\citenamefont {Behunin}, \citenamefont {Ou},\ and\ \citenamefont {Kieu}}]{behunin_spontaneous_2019}%
  \BibitemOpen
  \bibfield  {author} {\bibinfo {author} {\bibfnamefont {R.~O.}\ \bibnamefont {Behunin}}, \bibinfo {author} {\bibfnamefont {Y.-H.}\ \bibnamefont {Ou}}, \ and\ \bibinfo {author} {\bibfnamefont {K.}~\bibnamefont {Kieu}},\ }\href {\doibase 10.1103/PhysRevA.99.063826} {\bibfield  {journal} {\bibinfo  {journal} {Physical Review A}\ }\textbf {\bibinfo {volume} {99}},\ \bibinfo {pages} {063826} (\bibinfo {year} {2019})},\ \bibinfo {note} {publisher: American Physical Society}\BibitemShut {NoStop}%
\bibitem [{\citenamefont {Shin}\ \emph {et~al.}(2013)\citenamefont {Shin}, \citenamefont {Qiu}, \citenamefont {Jarecki}, \citenamefont {Cox}, \citenamefont {Olsson}, \citenamefont {Starbuck}, \citenamefont {Wang},\ and\ \citenamefont {Rakich}}]{shin_tailorable_2013}%
  \BibitemOpen
  \bibfield  {author} {\bibinfo {author} {\bibfnamefont {H.}~\bibnamefont {Shin}}, \bibinfo {author} {\bibfnamefont {W.}~\bibnamefont {Qiu}}, \bibinfo {author} {\bibfnamefont {R.}~\bibnamefont {Jarecki}}, \bibinfo {author} {\bibfnamefont {J.~A.}\ \bibnamefont {Cox}}, \bibinfo {author} {\bibfnamefont {R.~H.}\ \bibnamefont {Olsson}}, \bibinfo {author} {\bibfnamefont {A.}~\bibnamefont {Starbuck}}, \bibinfo {author} {\bibfnamefont {Z.}~\bibnamefont {Wang}}, \ and\ \bibinfo {author} {\bibfnamefont {P.~T.}\ \bibnamefont {Rakich}},\ }\href {\doibase 10.1038/ncomms2943} {\bibfield  {journal} {\bibinfo  {journal} {Nature Communications}\ }\textbf {\bibinfo {volume} {4}},\ \bibinfo {pages} {1944} (\bibinfo {year} {2013})},\ \bibinfo {note} {number: 1 Publisher: Nature Publishing Group}\BibitemShut {NoStop}%
\bibitem [{\citenamefont {Kittlaus}\ \emph {et~al.}(2016)\citenamefont {Kittlaus}, \citenamefont {Shin},\ and\ \citenamefont {Rakich}}]{kittlaus_large_2016}%
  \BibitemOpen
  \bibfield  {author} {\bibinfo {author} {\bibfnamefont {E.~A.}\ \bibnamefont {Kittlaus}}, \bibinfo {author} {\bibfnamefont {H.}~\bibnamefont {Shin}}, \ and\ \bibinfo {author} {\bibfnamefont {P.~T.}\ \bibnamefont {Rakich}},\ }\href {\doibase 10.1038/nphoton.2016.112} {\bibfield  {journal} {\bibinfo  {journal} {Nature Photonics}\ }\textbf {\bibinfo {volume} {10}},\ \bibinfo {pages} {463} (\bibinfo {year} {2016})},\ \bibinfo {note} {publisher: Nature Publishing Group}\BibitemShut {NoStop}%
\bibitem [{\citenamefont {Shin}\ \emph {et~al.}(2015)\citenamefont {Shin}, \citenamefont {Cox}, \citenamefont {Jarecki}, \citenamefont {Starbuck}, \citenamefont {Wang},\ and\ \citenamefont {Rakich}}]{shin_control_2015}%
  \BibitemOpen
  \bibfield  {author} {\bibinfo {author} {\bibfnamefont {H.}~\bibnamefont {Shin}}, \bibinfo {author} {\bibfnamefont {J.~A.}\ \bibnamefont {Cox}}, \bibinfo {author} {\bibfnamefont {R.}~\bibnamefont {Jarecki}}, \bibinfo {author} {\bibfnamefont {A.}~\bibnamefont {Starbuck}}, \bibinfo {author} {\bibfnamefont {Z.}~\bibnamefont {Wang}}, \ and\ \bibinfo {author} {\bibfnamefont {P.~T.}\ \bibnamefont {Rakich}},\ }\href {\doibase 10.1038/ncomms7427} {\bibfield  {journal} {\bibinfo  {journal} {Nature Communications}\ }\textbf {\bibinfo {volume} {6}},\ \bibinfo {pages} {6427} (\bibinfo {year} {2015})},\ \bibinfo {note} {publisher: Nature Publishing Group}\BibitemShut {NoStop}%
\bibitem [{\citenamefont {Zhang}\ \emph {et~al.}(2017)\citenamefont {Zhang}, \citenamefont {Wang}, \citenamefont {Cheng},\ and\ \citenamefont {Tsang}}]{zhang_forward_2017}%
  \BibitemOpen
  \bibfield  {author} {\bibinfo {author} {\bibfnamefont {Y.}~\bibnamefont {Zhang}}, \bibinfo {author} {\bibfnamefont {L.}~\bibnamefont {Wang}}, \bibinfo {author} {\bibfnamefont {Z.}~\bibnamefont {Cheng}}, \ and\ \bibinfo {author} {\bibfnamefont {H.~K.}\ \bibnamefont {Tsang}},\ }\href {\doibase 10.1063/1.4996367} {\bibfield  {journal} {\bibinfo  {journal} {Applied Physics Letters}\ }\textbf {\bibinfo {volume} {111}},\ \bibinfo {pages} {041104} (\bibinfo {year} {2017})}\BibitemShut {NoStop}%
\bibitem [{\citenamefont {Bahl}\ \emph {et~al.}(2013)\citenamefont {Bahl}, \citenamefont {Kim}, \citenamefont {Lee}, \citenamefont {Liu}, \citenamefont {Fan},\ and\ \citenamefont {Carmon}}]{bahl_brillouin_2013}%
  \BibitemOpen
  \bibfield  {author} {\bibinfo {author} {\bibfnamefont {G.}~\bibnamefont {Bahl}}, \bibinfo {author} {\bibfnamefont {K.~H.}\ \bibnamefont {Kim}}, \bibinfo {author} {\bibfnamefont {W.}~\bibnamefont {Lee}}, \bibinfo {author} {\bibfnamefont {J.}~\bibnamefont {Liu}}, \bibinfo {author} {\bibfnamefont {X.}~\bibnamefont {Fan}}, \ and\ \bibinfo {author} {\bibfnamefont {T.}~\bibnamefont {Carmon}},\ }\href {\doibase 10.1038/ncomms2994} {\bibfield  {journal} {\bibinfo  {journal} {Nature Communications}\ }\textbf {\bibinfo {volume} {4}},\ \bibinfo {pages} {1994} (\bibinfo {year} {2013})},\ \bibinfo {note} {publisher: Nature Publishing Group}\BibitemShut {NoStop}%
\bibitem [{\citenamefont {Bahl}\ \emph {et~al.}(2011)\citenamefont {Bahl}, \citenamefont {Zehnpfennig}, \citenamefont {Tomes},\ and\ \citenamefont {Carmon}}]{bahl_stimulated_2011}%
  \BibitemOpen
  \bibfield  {author} {\bibinfo {author} {\bibfnamefont {G.}~\bibnamefont {Bahl}}, \bibinfo {author} {\bibfnamefont {J.}~\bibnamefont {Zehnpfennig}}, \bibinfo {author} {\bibfnamefont {M.}~\bibnamefont {Tomes}}, \ and\ \bibinfo {author} {\bibfnamefont {T.}~\bibnamefont {Carmon}},\ }\href {\doibase 10.1038/ncomms1412} {\bibfield  {journal} {\bibinfo  {journal} {Nature Communications}\ }\textbf {\bibinfo {volume} {2}},\ \bibinfo {pages} {403} (\bibinfo {year} {2011})},\ \bibinfo {note} {publisher: Nature Publishing Group}\BibitemShut {NoStop}%
\bibitem [{\citenamefont {Yu}\ \emph {et~al.}(2022)\citenamefont {Yu}, \citenamefont {Shen}, \citenamefont {Yang}, \citenamefont {Qi}, \citenamefont {Jiang}, \citenamefont {Brambilla}, \citenamefont {Dong},\ and\ \citenamefont {Wang}}]{yu_investigation_2022}%
  \BibitemOpen
  \bibfield  {author} {\bibinfo {author} {\bibfnamefont {J.}~\bibnamefont {Yu}}, \bibinfo {author} {\bibfnamefont {Z.}~\bibnamefont {Shen}}, \bibinfo {author} {\bibfnamefont {Z.}~\bibnamefont {Yang}}, \bibinfo {author} {\bibfnamefont {S.}~\bibnamefont {Qi}}, \bibinfo {author} {\bibfnamefont {Y.}~\bibnamefont {Jiang}}, \bibinfo {author} {\bibfnamefont {G.}~\bibnamefont {Brambilla}}, \bibinfo {author} {\bibfnamefont {C.-H.}\ \bibnamefont {Dong}}, \ and\ \bibinfo {author} {\bibfnamefont {P.}~\bibnamefont {Wang}},\ }\href {\doibase 10.1109/JPHOT.2022.3145033} {\bibfield  {journal} {\bibinfo  {journal} {IEEE Photonics Journal}\ }\textbf {\bibinfo {volume} {14}},\ \bibinfo {pages} {1} (\bibinfo {year} {2022})},\ \bibinfo {note} {conference Name: IEEE Photonics Journal}\BibitemShut {NoStop}%
\bibitem [{\citenamefont {Dupays}\ and\ \citenamefont {Pain}(2023)}]{dupays_closed_2023}%
  \BibitemOpen
  \bibfield  {author} {\bibinfo {author} {\bibfnamefont {L.}~\bibnamefont {Dupays}}\ and\ \bibinfo {author} {\bibfnamefont {J.-C.}\ \bibnamefont {Pain}},\ }\href {\doibase 10.1088/1751-8121/acc8a3} {\bibfield  {journal} {\bibinfo  {journal} {Journal of Physics A: Mathematical and Theoretical}\ }\textbf {\bibinfo {volume} {56}},\ \bibinfo {pages} {255202} (\bibinfo {year} {2023})},\ \bibinfo {note} {arXiv:2107.01204 [math-ph]}\BibitemShut {NoStop}%
\bibitem [{\citenamefont {Meystre}(2021)}]{meystre_quantum_2021}%
  \BibitemOpen
  \bibfield  {author} {\bibinfo {author} {\bibfnamefont {P.}~\bibnamefont {Meystre}},\ }\href {\doibase 10.1007/978-3-030-76183-7} {{\selectlanguage {en}\emph {\bibinfo {title} {Quantum {Optics}: {Taming} the {Quantum}}}}},\ Graduate {Texts} in {Physics}\ (\bibinfo  {publisher} {Springer International Publishing},\ \bibinfo {address} {Cham},\ \bibinfo {year} {2021})\BibitemShut {NoStop}%
\bibitem [{\citenamefont {Lee}\ and\ \citenamefont {Agrawal}(2003)}]{lee_suppression_2003}%
  \BibitemOpen
  \bibfield  {author} {\bibinfo {author} {\bibfnamefont {H.}~\bibnamefont {Lee}}\ and\ \bibinfo {author} {\bibfnamefont {G.~P.}\ \bibnamefont {Agrawal}},\ }\href {\doibase 10.1364/OE.11.003467} {\bibfield  {journal} {\bibinfo  {journal} {Optics Express}\ }\textbf {\bibinfo {volume} {11}},\ \bibinfo {pages} {3467} (\bibinfo {year} {2003})},\ \bibinfo {note} {publisher: Optica Publishing Group}\BibitemShut {NoStop}%
\bibitem [{\citenamefont {Merklein}\ \emph {et~al.}(2015)\citenamefont {Merklein}, \citenamefont {Kabakova}, \citenamefont {Büttner}, \citenamefont {Choi}, \citenamefont {Luther-Davies}, \citenamefont {Madden},\ and\ \citenamefont {Eggleton}}]{merklein_enhancing_2015}%
  \BibitemOpen
  \bibfield  {author} {\bibinfo {author} {\bibfnamefont {M.}~\bibnamefont {Merklein}}, \bibinfo {author} {\bibfnamefont {I.~V.}\ \bibnamefont {Kabakova}}, \bibinfo {author} {\bibfnamefont {T.~F.~S.}\ \bibnamefont {Büttner}}, \bibinfo {author} {\bibfnamefont {D.-Y.}\ \bibnamefont {Choi}}, \bibinfo {author} {\bibfnamefont {B.}~\bibnamefont {Luther-Davies}}, \bibinfo {author} {\bibfnamefont {S.~J.}\ \bibnamefont {Madden}}, \ and\ \bibinfo {author} {\bibfnamefont {B.~J.}\ \bibnamefont {Eggleton}},\ }\href {\doibase 10.1038/ncomms7396} {\bibfield  {journal} {\bibinfo  {journal} {Nature Communications}\ }\textbf {\bibinfo {volume} {6}},\ \bibinfo {pages} {6396} (\bibinfo {year} {2015})},\ \bibinfo {note} {publisher: Nature Publishing Group}\BibitemShut {NoStop}%
\bibitem [{\citenamefont {Puckett}\ \emph {et~al.}(2019)\citenamefont {Puckett}, \citenamefont {Bose}, \citenamefont {Nelson},\ and\ \citenamefont {Blumenthal}}]{puckett_higher_2019}%
  \BibitemOpen
  \bibfield  {author} {\bibinfo {author} {\bibfnamefont {M.}~\bibnamefont {Puckett}}, \bibinfo {author} {\bibfnamefont {D.}~\bibnamefont {Bose}}, \bibinfo {author} {\bibfnamefont {K.}~\bibnamefont {Nelson}}, \ and\ \bibinfo {author} {\bibfnamefont {D.~J.}\ \bibnamefont {Blumenthal}},\ }in\ \href {\doibase 10.1364/CLEO_SI.2019.SM4O.1} {\emph {\bibinfo {booktitle} {2019 {Conference} on {Lasers} and {Electro}-{Optics} ({CLEO})}}}\ (\bibinfo {year} {2019})\ pp.\ \bibinfo {pages} {1--2},\ \bibinfo {note} {iSSN: 2160-8989}\BibitemShut {NoStop}%
\bibitem [{\citenamefont {Wang}\ \emph {et~al.}(2024)\citenamefont {Wang}, \citenamefont {Hu}, \citenamefont {Lao}, \citenamefont {Wang}, \citenamefont {Jin}, \citenamefont {Zhou}, \citenamefont {Lei}, \citenamefont {Wang}, \citenamefont {Liu}, \citenamefont {Yang},\ and\ \citenamefont {Li}}]{wang_taming_2024}%
  \BibitemOpen
  \bibfield  {author} {\bibinfo {author} {\bibfnamefont {M.}~\bibnamefont {Wang}}, \bibinfo {author} {\bibfnamefont {Z.-G.}\ \bibnamefont {Hu}}, \bibinfo {author} {\bibfnamefont {C.}~\bibnamefont {Lao}}, \bibinfo {author} {\bibfnamefont {Y.}~\bibnamefont {Wang}}, \bibinfo {author} {\bibfnamefont {X.}~\bibnamefont {Jin}}, \bibinfo {author} {\bibfnamefont {X.}~\bibnamefont {Zhou}}, \bibinfo {author} {\bibfnamefont {Y.}~\bibnamefont {Lei}}, \bibinfo {author} {\bibfnamefont {Z.}~\bibnamefont {Wang}}, \bibinfo {author} {\bibfnamefont {W.}~\bibnamefont {Liu}}, \bibinfo {author} {\bibfnamefont {Q.-F.}\ \bibnamefont {Yang}}, \ and\ \bibinfo {author} {\bibfnamefont {B.-B.}\ \bibnamefont {Li}},\ }\href {\doibase 10.1103/PhysRevX.14.011056} {\bibfield  {journal} {\bibinfo  {journal} {Physical Review X}\ }\textbf {\bibinfo {volume} {14}},\ \bibinfo {pages} {011056} (\bibinfo {year} {2024})},\ \bibinfo {note} {publisher: American Physical Society}\BibitemShut {NoStop}%
\end{thebibliography}
